\documentstyle[12pt]{article}

\newcommand{\bb}{\begin{equation}}
\newcommand{\ee}{\end{equation}}
\newcommand{\pk}{\partial_{\mu_1 ...\mu_k}}
\newcommand{\pss}{\partial_{\mu_1 ...\mu_s}}
\newcommand{\dal}{\Delta}

\newcommand{\pj}{\partial_{\mu_1 ...\mu_j}}

\begin{document}

{}~\\
{}~\\
\begin{center}
{\huge {\bf Spacetime locality of the antifield formalism :
 a ``mise au point"\footnote {Based on the author's
contribution to the
Proceedings of the second meeting
on constrained Hamiltonian systems, Montepulciano,
Italy, June 28-July 2,
1993.}}}\\
\vspace{2.5cm}
{\large Marc Henneaux$^{a,b}$}, \\\vspace{1.5cm}
{}~$^a$ {\em Facult\'e des Sciences, Universit\'e
Libre de Bruxelles, Campus Plaine C.P. 231, B-1050
Bruxelles, Belgium.} \\
{}~$^b${\em Centro de Estudios Cient\'{\i}ficos de Santiago,
Casilla 16443, Santiago 9, Chile} \\
\vspace{2cm}
\end{center}

\begin{abstract}

\noindent
Some general techniques and theorems on the spacetime
locality of the antifield
formalism are illustrated in the familiar cases of
the free scalar field, electromagnetism and Yang-Mills theory.
The analysis explicitly shows that recent criticisms of the usual
approach to dealing with locality
are ill-founded.

\end{abstract}

\vfill
\break

\newpage
\section{Introduction}

The antifield-antibracket formalism appears
to be the most elegant and powerful
method for quantizing gauge theories \cite{Bat,HT}.
In that formalism, new variables, called
the ``antifields", play a central role.
As it has been shown in \cite{FHST,FH}, the rationale for
introducing the antifields is that these provide a resolution
of the algebra of functionals of on-shell field configurations.
Namely, the antifields are there to implement the equations
of motion when one passes to the BRST cohomology.
The resolution associated with the antifields is called
``Koszul-Tate" resolution, because it is patterned after a
construction due to Koszul \cite{Kos},
supplemented, when the equations
of motion are not independent, by the introduction of further
variables killing unwanted homology along lines due to Tate
\cite{Tate}. [We assume
some familiarity with the general ideas
of the antifield formalism;
we refer to \cite{HT} for a detailed exposition].

The analysis presented in \cite{FH} did not address
the question of the spacetime locality of the construction.
A few years ago, that question has been investigated and
completely solved
\cite{Hen} (see also \cite{HT}, chapters
12 and 17).  The analysis of \cite{Hen} has been critized
or ignored, however, on the (incorrect) grounds that the
locality conditions assumed in it would not apply to the
usual gauge theories of physical interest \cite{VP}.  The
purpose of this paper is to make it clear how the approach of
\cite{Hen} works and does indeed solve the issue of locality by
illustrating it in the familiar cases of the Klein-Gordon field,
the electromagnetic field and the Yang-Mills field.

We shall analyse only the specific question of locality of the
Koszul-Tate complex.  The reference \cite{HT} contains a
discussion as to why this complex is so useful in the
quantization of gauge systems.

\section{Definitions}

Consider a field theory with field variables $\phi^i$.
We shall deal with both local functionals and local functions of
$\phi^i$.  Local functions are functions of $\phi^i$ and a finite
number of their derivatives, which may also involve the spacetime
coordinates explicitly.  So, a local function is
given by
\bb
f(x^\mu, \phi^i, \partial_\mu \phi^i, ..., \pk \phi^i).
\ee
Local functionals are integrals of local functions.
Hence,
\bb
F[\phi^i] = \int f(x^\mu, \phi^i, \partial_\mu \phi^i, ..., \pk \phi^i) d^n x
\ee
is a local functional.

The appropriate way to deal with local functions
is well known and has been used quite a lot in the algebraic
study of anomalies.  The corresponding
mathematical framework
is the one of jet bundle theory (see e.g. \cite{And,Bry}).
However, in order to keep the discussion simple, we shall not
adopt here the jet bundle terminology.  This is permissible
because we shall assume that spacetime is $R^n$, so that there
are no global subtleties.

Let $V^0$ be the space with coordinates ($x,\phi^i$).  More
generally, let $V^k$ be the space with coordinates ($x,\phi^i,
\partial_\mu \phi^i, ..., \pk \phi^i$).  If $f$ is a smooth local
function, then there exists $k$ such that $f \in C^{\infty}(V^k)$.
For this reason, the $V^k$'s are the natural spaces in which
to analyze locality.  These spaces arose
first in the geometric study of differential equations, which can
naturally be regarded as representing surfaces in the $V^k$'s.
In that context, the spaces $V^k$ are called $k$-th jet bundles and
are denoted by $J^k(E)$.

We stress that the jet bundle spaces are
quite familiar not only in mathematics but also in
physics since these are the spaces in which the Lagrangians of
local field theories
live.  These spaces are finite dimensional for each $k$.
For this reason, all the standard algebraic
tools of the antifield formalism
(contracting homotopy, counting operators, recursive introduction
of the antifields of antifields by successive killing of
unwanted cohomology, model for the exterior derivative
along the gauge orbits, antibracket cohomology, role of
zeroth order terms - see \cite{HT}) are available
in the jet bundle spaces without functional
complications.

In order to discuss local functionals, it is useful to consider
the algebra $A_k \equiv C^{\infty}(V^k) \otimes \bigwedge [dx^\mu]$
of exterior forms on $R^n$ with coefficients that are functions
on $V^k$,
\bb
\omega \in A_k  \Leftrightarrow \omega =
\Sigma \; \omega_{\nu_1 ... \nu_j}(x,\phi^i,
\partial_\mu \phi^i, ..., \pk \phi^i)
\, dx^{\nu_1}\wedge...\wedge dx^{\nu_j}
\ee
One can define a differential $d : A_k \rightarrow A_{k+1}$
as follows,
\bb
d \omega = \Sigma \; d \omega_{\mu_1 ... \mu_j}\wedge
\, dx^{\mu_1}\wedge...\wedge dx^{\mu_j}
\ee
where $d$ acting on a function $f \in A_k$ is defined by
\bb
df = \frac {\partial^T f}{\partial x^\mu}\, dx^\mu,
\ee
\bb
\frac {\partial^T f}{\partial x^\mu}
\equiv \frac {\partial f}{\partial x^\mu} +
\frac {\partial f}{\partial\phi^i} \partial_\mu \phi^i +
... + \frac {\partial f}{\partial(\pk)\phi^i}\, \partial_{\mu_1 ... \mu_k \mu}
\phi^i.
\ee
One crucial property of $d$ is that
\bb
\label{Stokes}
\int d \omega = 0
\ee
(we assume here and throughout that the boundary conditions
are such that the surface terms appearing in the equations
vanish. If not, one must carefully keep track of the relevant
surface integrals).

Conversely let $\rho$ be a $n$-form
such that $\int \rho = 0$ for all field configurations.
Then $\rho = d \omega$ (see e.g. \cite{HT}).  Accordingly, two
local functions determine the same local functional if and only
if they differ by a $d$-exact term.  For that reason, one can,
following Gel'fand and Dorfman \cite{Gel}, identify local functionals
with the quotient space $H^n(d)$ of local $n$-forms (which
are automatically closed) modulo exact ones.

The Lagrangian ${\cal L}(\phi^i,\partial_\mu \phi^i,...,
\pss\phi^i)$ of the theory is a smooth function on
$V^s$.  The
equations of motion{\footnote {From now on, we shall drop the
suffix $T$ on $\partial^T_\mu$: $\partial_\mu$ always stands for
$\partial^T_\mu$.}}
\bb
\frac {\delta {\cal L}}{\delta \phi^i} \equiv
\frac {\partial {\cal L}}{\partial \phi^i} -
\partial_\mu \frac {\partial {\cal L}}{\partial (\partial_\mu \phi^i)}
+ ...+ (-1)^s \pss \frac {\partial {\cal L}}{\partial (\pss \phi^i)},
\ee
together with their derivatives $\partial_\mu
(\delta {\cal L}/\delta \phi^i)
= 0$, $\partial_{\mu_1 \mu_2} (\delta {\cal L}/\delta \phi^i) = 0$ ...
determine surfaces $\Sigma_k$ in $V^k$.  For a fixed $k$, only a finite
number of equations are relevant.  The surfaces $\Sigma_k$ are called
``stationary surfaces".

In the antifield formalism, the algebra $C^{\infty}(\Sigma_k)$ of
smooth functions on $\Sigma_k$ plays an important role because
it is related to the observables \cite{HT}.  The Koszul-Tate
construction provides a resolution of $C^{\infty}(\Sigma_k)$ for
each $k$.  The idea is to view $C^{\infty}(\Sigma_k)$ as the
quotient algebra $C^{\infty}(V^k)/{\cal N}_k$, where ${\cal N}_k$
is the ideal of functions of $C^{\infty}(V^k)$ that vanish on
$\Sigma_k$. The Koszul-Tate differential is such that the elements
of ${\cal N}_k$ are exact, i.e., are pure boundaries.

\section{The Koszul-Tate differential for the
massless scalar field}

To illustrate the construction, we consider first the massless
Klein-Gordon theory.  One has a single scalar field $\phi$ with
Lagrangian
\bb
{\cal L} = - \frac{1}{2} \partial_\mu \phi \partial^\mu \phi
\ee
The equations of motion are
\bb
\dal \phi \equiv \partial_\mu\partial^\mu \phi = 0.
\ee
In $V^0$, the equations of motion imply no relation
and $\Sigma_0$ is empty: two functions
$f$ and $g$ in $V^0$ coincide ``on-shell" (i.e., when the
equations of motion hold) if and only if they are identical.
Similarly, there is no relation in
$V^1$.  One has to go to $V^2$ to see the first effect of the
equations of motion, which restrict the second derivatives
of $\phi$.  The surface $\Sigma_2$ is defined by $\dal \phi = 0$
in $V^2$.  Then, in $V^3$, $\Sigma_3$ is the surface $\dal \phi = 0$,
$\partial_\mu \dal \phi = 0$. More generally, the surface $\Sigma_k$ in
$V^k$ is defined by the equations
\bb
\label{KG}
\Sigma_k : \dal \phi = 0, ..., \dal \partial_{\mu_1}...
\partial_{\mu_{k-2}} \phi = 0.
\ee

The equations of motion (\ref{KG}) are independent in $V^k$.
This is most easily seen by introducing a new coordinate system
in $V^k$, which has the left hand side of the equations (\ref{KG})
as independent coordinates.  One such coordinate system is given by
\bb
\label{coord}
\phi,\, \partial_\mu \phi,\, \partial_{m_1 m_2} \phi, \,
\partial_{m_1 0}\phi, \dal \phi, ...,
\partial_{m_1...m_{k-3}m_{k}} \phi,\, \partial_{m_1...m_{k-1}0} \phi,
\,\partial_{\mu_1 ... \mu_{k-2}} \dal \phi.
\ee
One can easily verify that any function $f$ on $V^k$ that vanishes on
$\Sigma_k$ $(f \approx 0$) takes the form,
\bb
\label{weak}
f \approx 0 \Leftrightarrow
f = h \dal \phi + h^\mu \partial_\mu \dal \phi + ...
+ h^{\mu_1 ... \mu_{k-2}} \dal \partial_{\mu_1}...\partial_{\mu_{k-2}}
\phi
\ee
where the $h$'s are functions on $V^k$
(see for instance \cite{HT}, chapter 1 with $\phi_m = 0$
replaced by (\ref{KG})).

In order to construct a resolution of $C^\infty(\Sigma_k)$,
one introduces one independent odd generator for each (independent)
equation (\ref{KG}).  That is, one considers the differential
algebra $C^\infty(V^k) \otimes \bigwedge [\phi^*, \partial_\mu\phi^*,
..., \partial_{\mu_1}...\partial_{\mu_{k-2}}\phi^*]$ with
differential
\bb
\label{exact1}
\delta \phi = 0, \delta \phi^* = \dal \phi,
\ee
extended to the derivatives of the field and ``antifield" $\phi^*$
so as to commute with $\partial_\mu$,
\bb
\label{exact2}
\delta \pj \phi = 0, \delta \pj \phi^* = \pj \dal \phi.
\ee
One defines also the antighost number through
\bb
antigh(\phi) = 0, \; antigh(\phi^*) = 1.
\ee
By (\ref{exact1}), (\ref{exact2}), every equation of motion is
$\delta$-exact and so, is identified with zero when
one passes to the $\delta$-homology.  More
precisely, standard arguments from homological algebra show that
\bb
\label{acyc1}
H_0(\delta) = C^\infty(\Sigma_k),\;\,  H_j(\delta) = 0 \hbox{ for }
 \, j \neq 0.
\ee

This result may be derived by observing that the coordinates of
$C^\infty(V^k) \otimes \bigwedge [\phi^*, \partial_\mu\phi^*,
..., \partial_{\mu_1}...\partial_{\mu_{k-2}}\phi^*]$
 split
into three groups $(x_i, z_\alpha,J{\cal P}_\alpha)$ such
that $\delta$ takes the form
\bb
\delta x_i = 0,\; \delta {\cal P}_\alpha = z_\alpha, \;
\delta z_\alpha = 0
\ee
or equivalently
\bb
\delta = z_\alpha {\frac {\partial}{\partial {\cal P}_\alpha}}.
\ee
Explicitly, the coordinates $x_i$ stand
for the field $\phi$ and its derivatives with at most one $\partial_0$, the
$z_\alpha$ stand for $\dal \phi$ and its derivatives, while the
${\cal P}_\alpha$ stand for $\phi^*$ and its derivatives.  A contracting
homotopy may be defined through
\bb
\sigma x_i = 0,\, \sigma {\cal P}_\alpha = 0,\, \sigma z_\alpha
= {\cal P}_\alpha \, \Leftrightarrow  \sigma = {\cal P}_\alpha {\frac
{\partial}{\partial z_\alpha}},
\ee
i.e.,
\bb
\label{contr1}
\sigma = \phi^* \frac {\partial}{\partial (\dal \phi)} +
\partial_\mu \phi^* \frac {\partial}{\partial (\partial_\mu\dal \phi)}
+...+ \partial_{\mu_1}...\partial_{\mu_{k-2}}\phi^*
\frac {\partial}{\partial (\partial_{\mu_1...\mu_{k-2}} \dal \phi)}
\ee
where the derivatives with respect to $\pj \dal \phi$
are computed in the coordinates (\ref{coord}) of $V^k$.
One has
\bb
\label{sigmaN}
\sigma \delta + \delta \sigma = N
\ee
where $N$
\bb
\label{defN}
N = {\cal P}_\alpha {\frac {\partial}{\partial {\cal P}_\alpha}} +
z_\alpha {\frac {\partial}{\partial z_\alpha}}
\ee
is the operator counting the number of ${\cal P}_\alpha$ and
$z_\alpha$.  The relation (\ref{sigmaN}) crucially uses the
derivation property of $\partial / \partial z_\alpha$.
It follows from (\ref{sigmaN}) and (\ref{defN})
that ${\cal P}_\alpha$
and $z_\alpha$ drop from the homology of $\delta$
(``they belong to the contractible part of the complex"), which is
given by the functions of $x_i$ (\cite{HT}, sections 8.3.2 and
9.A.2.  The $G_a$'s there play the role of the equations of motion
here).  Since the functions of $x_i$ are the functions on $\Sigma_k$ and
have antighost number equal to zero, formula (\ref{acyc1}) is
established.

The argument is valid for any $k$, i.e. for any local function involving
the derivatives of the field and antifield up to an arbitrarily
high (but finite) order.  One sometimes summarize (\ref{acyc1}) by
saying that $\delta$ is acyclic in the space of local functions.

It should be noted that even though
covariant-looking, the contracting homotopy (\ref{contr1})
is not covariant.  For instance, one finds
\bb
\sigma (\partial_\mu \partial_\nu \phi) = \delta_{\mu 0}  \delta_{\nu 0}
\phi^*.
\ee
Nevertherless, one can show that the homology of $\delta$ in the
algebra of Lorentz invariant functions is trivial for positive
$k$; that is, if $\delta f = 0$ and $antigh(f) = k \neq 0$, where
$f$ is Lorentz invariant, then $f = \delta g$ where $g$ may
also be taken to be Lorentz invariant. This can be proved either
by redefining the homotopy, or equivalently, by following the methods of
\cite{Henn}, theorem 2.

We close this section by a few remarks concerning incorrect statements
that have been made in the literature.

\noindent
1. First, it
should be stressed that $f \approx 0$ does not imply $f = h \dal \phi$
with $h$ a local function.  Rather,
$f$ may also involve the derivatives of $\dal \phi$, i.e.,
one has the full expansion (\ref{weak}).

\noindent
2. The homotopy $\sigma$ given by (\ref{contr1})
is well defined everywhere because
the equations of motion are
simple.  For more general theories, however,
a globally defined homotopy constructed along the above lines
may just simply not exist.  This is because obstructions
for defining the derivation $\partial/
\partial (\delta {\cal L}/\delta \phi^i)$ may be present (one
needs to tell what is kept fixed when differentiating with respect
to $\delta {\cal L}/\delta \phi^i$).
Attempts for using a formula similar to (\ref{contr1}) would then
necessarily fail.  This would show up in non convergence of
power series, etc., which must be handled carefully.
This difficulty has
been overlooked in \cite{VP}.
One way to handle correctly this problem
 is to introduce partitions of unity,
as in \cite{HT}, appendix 9A.

To make this point clear, consider the
Lagrangian
$L = L(q)$
where the function $h(q) \equiv dL/dq$ is such that (i) $h(q) = -1$
for $q \leq -1$; (ii) $h(q) = 1$ for $q \geq 1$; and (iii) $h(q)$ interpolates
in a smooth way from $-1$ to $+1$ between
$-1$ to $+1$ and vanishes only at the origin where
$h'(0) = 1$.  It is clear that it is impossible to define $df/dh$ for
all functions $f$'s (with $d/dh$ a derivation)
since this would imply in particular that
$dq/dh$ is well-defined and such that
$(dq/dh)(dh/dq) = 1$, in contradiction with $dh/dq = 0$ for
$q \leq -1$ or $q \geq 1$.
It turns out not
to be necessary, however, to define
$df/dh$ in the open sets where $h \not= 0$.  Indeed, in those sets
(``of type V" according to \cite{HT}), any $\delta$-closed
function $f$ is trivially $\delta$-exact, $f = \delta (q^* f/h)$.  The
proof of acyclicity of $\delta$ proceeds
by patching the $V$-sets with an open set covering the origin by
means of a partition of unity.

One may also construct polynomial counterexamples.
For instance, the Lagrangian
\bb
L(q) = \frac{1}{4} q^4 + \frac{5}{3} q^3 + \frac{1}{2} q^2 + 5 q
\ee
for a real variable $q$ leads to
the equation of motion $h(q) \equiv dL/dq = (q+5) (q^2 + 1) = 0$,
whose sole solution is $q=-5$.  The equation of motion is regular
($h'(q) \not= 0$ on-shell), but yet, one cannot define $dq/dh$
everywhere since $dh/dq$ has two real roots.  One may build
other counteramples based on a non trivial topology of the
stationary surface.

\vspace{.2cm}
\noindent
3.  In the same way, given a function on the stationary
surface, one can extend it off the constraint surface so
that it vanishes outside a tubular neighbourghood of $\Sigma_k$.
That is, any element of $C^\infty(\Sigma_k)$ has a representative
in $C^\infty(V_k)$ that vanishes sufficiently away from $\Sigma_k$.  The
use of such representatives would clearly make the power expansions
considered in \cite{VP} problematical far from $\Sigma_k$.

\section{The Koszul-Tate differential for
the electromagnetic field}

We now turn to the electromagnetic case.  The equations of motion are
\bb
{\cal L}^\rho \equiv \frac {\delta {\cal L}}{\delta A_\rho} =
\partial_\mu F^{\mu \rho} = 0
\ee
and define a surface in $V^2$.  The new feature compared with the
previous situation is that the derived equations
\bb
\partial_\mu {\cal L}^\rho  = 0, \partial_{\mu_1\mu_2} {\cal L}^\rho
= 0, ...
\ee
in $V^3$, $V^4$, ... are no longer independent.  Because of the
gauge invariance of the electromagnetic field
Lagrangian, one has rather (identically)
\bb
\partial_\rho {\cal L}^\rho  \equiv 0, \partial_{\mu_1}
(\partial_\rho {\cal L}^\rho ) \equiv 0 ...
\ee
(for any field configuration).
For that reason, one needs ``antifields of anti- \\
fields"J\cite{FH,HT}.

We start with $V^2$.  There are clearly no relations among the equations
${\cal L}^\rho  = 0$ in $V^2$ since
one can solve these equations for $n$ of the
coordinates in $V^2$ (we work in $n$ dimensions).  Namely, one
can solve ${\cal L}^k = 0$ for $\partial_{00} A_k$ and ${\cal L}^0 = 0$
for $\partial_{11} A_0$ (say).  Hence, if one defines in
$C^\infty (V^2) \otimes \bigwedge (A^{*\mu})$ the differential
\bb
\label{KT1}
\delta A_\mu = 0, \delta \partial_\rho A_\mu = 0,
\delta \partial_{\rho \sigma} A_\mu = 0,
\delta A^{*\mu} = \partial_\nu F^{\nu \mu}
\ee
one gets that $H_k (\delta) = 0$ for $k \not= 0$
and $H_0 (\delta) = C^\infty
(\Sigma_2)$.  To verify this statement,
one repeats the argument of the previous section
and splits the variables of the complex in three groups.
The coordinates $A_\mu$, $\partial_\rho A_\mu$,
$\partial_{\rho \sigma} A^k$ ($(\rho, \sigma) \not= (0,0)$) and
$\partial_{\rho \sigma} A^0$  ($(\rho, \sigma) \not= (1,1)$)  are
of the $x_i$-type, the coordinates ${\cal L}^\rho$ are of the
$z_\alpha$-type, while the $A^{*\mu}$ are of the ${\cal P}_\alpha$-type.  The
appropriate contracting homotopy in
$C^\infty (V^2) \otimes \bigwedge (A^{*\mu})$ reads
\bb
\label{homoto1}
\sigma = A^{*\mu} \frac{\partial}{\partial{\cal L}^\mu}.
\ee
Thus, only the variables not constrained
by the equations of motion, namely, $A_\mu$, $\partial_\rho A_\mu$,
$\partial_{\rho \sigma} A^k$ ($(\rho, \sigma) \not= (0,0)$) and
$\partial_{\rho \sigma} A^0$  ($(\rho, \sigma) \not= (1,1)$) remain
in homology.  The other variables drop out.

Turn now to $C^\infty (V^3) \otimes \bigwedge (A^{*\mu},
\partial_{\rho} A^{*\mu})$, with differential $\delta$ (\ref{KT1}) extended
to the derivatives so that
\bb
\delta \partial_\mu = \partial_\mu \delta
\ee
i.e.,
\bb
\label{KT2}
\delta \partial_{\rho \sigma \alpha} A_{\mu} = 0,\;\;
\delta \partial_\rho A^{*J\mu} = \partial_\rho (\partial_\nu F^{\nu \mu})
\ee
The equations $\partial_\nu F^{\nu \mu} = 0$ and $\partial_\sigma
\partial_\nu F^{\nu \mu} = 0$ are {\it not} independent in $V^3$ since they
are subject to the (single) condition $\partial_\rho {\cal L}^\rho = 0$.
There are no other identity in $V^3$ because one can solve $n^2+n - 1$ of
the $n^2 + n$ equations ${\cal L}^\rho = 0$, $\partial_\mu {\cal L}^\rho = 0$
for $n^2+n - 1$ independent variables, namely $\partial_{00} A_k$ (from
${\cal L}^k = 0$), $\partial_{11} A_0$ (from ${\cal L}^0 = 0$),
$\partial_{\rho 0 0} A_k$ (from $\partial_\rho {\cal L}^k = 0$)
and $\partial_{s 11} A_0$ (from $\partial_s
{\cal L}^0 = 0$).  The derivative $\partial_{0 1 1}A^0$ cannot be
determined from $\partial_0
{\cal L}^0 = 0$, which is not an independent equation
($\partial_0 {\cal L}^0 = - \partial_k {\cal L}^k $).  Hence, in
$V^3$, there are $n^2+n -1$ independent equations and $1$
dependent one.

Because the equations of motion in $V^3$ are not independent, there is
one non trivial cycle at antighost number $1$, namely $\partial_\rho
A^{*\rho}$.  Thus, $H_1(\delta) \not= 0$ in
$C^\infty (V^3) \otimes \bigwedge (A^{*\mu},
\partial_{\rho} A^{*\mu})$.  In order to achieve
acyclicity of the Koszul-Tate differential, one needs
to introduce one further even variable, denoted by $C^*$  and
called ``antifield of antifield" \cite{HT}, with grading
\bb
antigh C^* = 2.
\ee
This new variable must kill the non trivial cycle $\partial_\rho
A^{*\rho}$ in homology, so that one defines
\bb
\label{KT3}
\delta C^* = \partial_\rho A^{*\rho}.
\ee
Once $C^*$ is introduced, one can redefine the variables of
the differential complex
$C^\infty (V^3) \otimes {\it C}[A^{*\mu},
\partial_{\rho} A^{*\mu}, C^*]$
in such a way that $\delta$ takes again the characteristic form{\footnote
{From now on, we shall use the notation ${\it C}[A^{*\mu},
\partial_{\rho} A^{*\mu}, C^*]$ for the algebra $\bigwedge (A^{*\mu},
\partial_{\rho} A^{*\mu}) \otimes {\it R} [C^*]$.  The symmetry properties
are taken care of by the gradings of $A^{*\mu}$ (odd) and $C^*$ (even).}}
\bb
\delta x_i = 0, \;
\delta {\cal P}_\alpha = z_\alpha	, \;\delta z_\alpha = 0,
\ee
which makes manifest that $H_*(\delta) = C^\infty(x_i)$.
The variables $x_i$ have antighost number zero and parametrize
$\Sigma_3$.  They are explicitly given by $A_\mu$,
$\partial_\rho A_\mu$,
$\partial_{\rho \sigma} A_k$ ($(\rho, \sigma) \not= (0,0)$),
$\partial_{\rho \sigma} A_0$  ($(\rho, \sigma) \not= (1,1)$),
$\partial_{\rho \sigma \nu} A_k$ (with at most one time derivative)
and $\partial_{\rho \sigma \nu} A_0$  (with $(\rho, \sigma, \nu)
\not= (k,1,1)$ even up to a permutation).  The variables
${\cal P}_\alpha$ are $A^{*\mu}, \partial_\alpha A^{* k},
\partial_k A^{*0}$ {\it and} $C^*$.  The variables
$z_ \alpha$ are the left hand sides of the
equations of motion ${\cal L}^\rho, \partial_\alpha {\cal L}^k,
\partial_k {\cal L}^0$ {\it and} $\partial_\rho A^{*\rho}$.

The same pattern goes on with the higher order derivatives.  In
$C^\infty (V^k) \otimes {\it C}[A^{*\mu},
\partial_{\rho} A^{*\mu},..., \partial_{\rho_1...\rho_{k-2}} A^{*\mu},
 C^*,..., \partial_{\rho_1...\rho_{k-3}} C^*]$,
one may introduce new coordinates as follows:

\noindent
(i) Coordinates of $x_i$-type : $A_k$ and its derivatives with
at most one $\partial_0$; $A_0$ and its derivatives except
$\partial_{s_1 s_2 ...s_m} A_0$ with at least two $\partial_1$.
These variables parametrize $\Sigma_k$.

\noindent
(ii) Coordinates of $z_\alpha$-type: ${\cal L}^k$ and its
derivatives; ${\cal L}^0$ and its spatial derivatives;
$\partial_\rho A^{* \rho}$ and its derivatives.

\noindent
(iii) Coordinates of ${\cal P}_\alpha$-type : $A^{*k}$ and
its derivatives; $A^{*0}$ and its spatial derivatives; $C^*$
and its derivatives.

Thus, again, $H_0(\delta) = C^\infty(V^k)$ and $H_m(\delta)
= 0$, $m \not= 0$.  The contracting homotopy
has the standard form
\bb
\sigma = {\cal P}_\alpha {\frac {\partial}{\partial z_\alpha}}\; ,
\ee
where the sum runs over all the $z_\alpha$'s.
At each stage, one can separate the equations
${\cal L}^\rho = 0$ and their derivatives into
independent ones and dependent ones {\it without going
out of the spaces $V^k$, i.e., in a manner compatible
with spacetime locality}.  Statements to the contrary
\cite{VP} are thus wrong.

It is true that the dependent equations
at order $k+1$ are not just the derivatives of the dependent
equations at order $k$.  One cannot separate the $n$ equations
${\cal L}^\rho = 0$ into two groups, so that the independent
(respectively, dependent) equations would simply be all the
derivatives of the equations of the first (respectively second)
group. To achieve this property, one would have to make a non
local split.  But a split with this property is not necessary
once one formulates the
problem in terms of the standard spaces $V^k$ of jet bundle theory, as
appropriate for dealing with locality.

Similarly, although we have not done it,  one could define a
Lorentz-invariant homotopy by decomposing the
derivatives of the fields along the irreducible
representations of the Lorentz group.  Hence, acyclicity of the
Koszul-Tate differential also holds in the algebra of Lorentz-invariant
local functions.  This same result can equivalently
be established along the lines of
\cite{Henn}.

\section{The Koszul-Tate differential for
the Yang-Mills field}

The Yang-Mills case can be treated in the same manner.  This is
because the terms with the highest (second) order derivatives of the
gauge potential in the Yang-Mills equations of motion are exactly
the same as in the Abelian case.  Hence, the change of variables
such that the left hand sides of the equations of motion and their
derivatives are new coordinates is still permissible,
and one can proceed as above.

For instance, in $V^2$, one would take as new variables
$A^a_\mu, \partial_\rho A^a_\mu, \partial_{\rho \sigma} A^a_k$
($(\rho, \sigma) \not= (0,0)$), $\partial_{\rho \sigma} A^a_0$,
($(\rho, \sigma) \not= (1,1)$) and ${\cal L}^\mu _a$.  The
expression of $\partial_{00} A^a_k$ in terms of ${\cal L}^k _a$ is the
same as in the abelian case up to terms containing lower order derivatives
(which are independent coordinates in the
previous space $V^2$).  A similar analysis
holds for higher order derivatives.

We leave it to the reader to check also that an analogous
derivation can be performed for p-form gauge fields.  The
only difference is that one needs this time
more antifields for antifields because the reducibility equations
are not independent.

\section{Acyclicity of Koszul-Tate differential and local
functionals}

The above sections establish the acyclicity of $\delta$ in
the space of local functions.  Does this property also hold in the
space of local functionals?  That is, if $f$ is a $n$-form such that
\bb
\delta \int f = 0, \; antigh f \geq 1
\ee
does one have
\bb
\int f = \delta \int g
\ee
for some $n$-form $g$? [$f$ and $g$ are $n$-forms
with coefficients that are local functions].  Equivalently,
in terms of the integrands, does
\bb
\label{f1}
\delta f = dj, \;  antigh f \geq 1
\ee
imply
\bb
\label{f2}
f = \delta g + dk
\ee
for some $n$-form $g$ and $n-1$-form $k$?  The presence of the $d$-exact
terms in (\ref{f1}), (\ref{f2}) follows from (\ref{Stokes}) and {\it
must be taken into account}.  Failure to do
so would be incorrect.  The extra $d$-terms  in
(\ref{f1}) and (\ref{f2}) show that the relevant cohomology
when dealing with local functionals is the cohomology of $\delta$
modulo $d$ in the space of local $n$-forms.
The corresponding cohomological
spaces are  denoted $H_k(\delta /d)$.

As pointed out in \cite{Hen}, the answer to
this question is in general negative.  Constants of the motion define
non trivial solutions of $H_1(\delta /d)$.  Indeed,
the equation $\delta f + dj$ with $antigh f = -1$
and $antigh j = 0$ defines a conserved current $j$.  If
$f$ is trivial (of the form (\ref{f2})), then $j$ is a trivial
conserved current ($j = - \delta k + dm$).  Since there exist
in general non
trivial conserved currents, $H_1(\delta /d)$ is not empty.

However, if $f$ involves the ghosts{\footnote
{How the ghosts are introduced may be found
for example in \cite{HT}.  The
ghosts will be denoted by $C^\alpha$ and are annihilated
by the differential $\delta$.  Once the ghosts are introduced,
the cohomology of $\delta$ is given by $C^\infty (\Sigma_k)
\otimes \bigwedge (C^\alpha, \partial_\rho C^\alpha ...)$.}}
 - which is the case
encountered in homological perturbation theory -, then
(\ref{f1}) does imply (\ref{f2}).  To see this, consider
first the case where $f$ is linear in the ghosts.  By
making integrations by parts if necessary, one can assume that
$f$ does not involve the derivatives of the $C^\alpha$,
\bb
f = \lambda_\alpha C^\alpha, \; antigh \lambda_\alpha = 0.
\ee
Then, $\delta f = \delta (\lambda_\alpha) C^\alpha$.  If $\delta f =
dj$, then  $\delta f$ {\it and} $dj$ must separately vanish because
$dj$ would otherwise necessarily involve derivatives of the ghosts.
Thus $\delta \lambda_\alpha = 0$, which implies $\lambda_\alpha =
\delta \mu_\alpha$ since $H_k(\delta) = 0$ in the space of
local functions.  Consequently, $f = (\delta \mu_\alpha) C^\alpha =
\delta (\mu_\alpha C^\alpha)$, which is the sought-for result.
How to formalize the argument so that it applies also to forms
$f$ that are non linear in the ghosts is done in \cite{Hen}.  Thus, acyclicity
of $\delta$ holds in the space of local functionals involving
both the antifields and the ghosts.

\section{Conclusion}

We have illustrated in this paper how to handle locality
in the case of the antifield-antibracket formalism for gauge
field theories.  The tools involve both standard homological algebraic
techniques applied to finitely generated algebras and ideas from
jet bundle theory.  We have shown in particular how the equations of motion
for electromagnetism and Yang-Mills theory can split into independent
and dependent ones in the ``jet bundle" spaces $V^k$.
The tools illustrated here have been used recently
to prove a long-standing conjecture on the renormalization
of Yang-Mills models \cite{BH}.

We close this letter with two observations :

\noindent
(i)  The method of homological
perturbation theory is quite general and does not depend
on the precise form of the differential algebra on which
the derivations act, provided these derivations fulfill
the properties explained in \cite{HT} (chapter 8).  Thus, one may
modify the algebra of local functions by imposing restrictions
if one wishes to do so.  For instance,
the well-known theorem
that a BRST cohomological class is determined by
its component of order zero in the antifields
 is quite standard and follows from
the general principles of homological perturbation theory
(see again \cite{HT}, chapter 8, proof of main
theorem and section 8.4.4).

\noindent
(ii)  Similarly, one may consider field theories for which
the equations of motion are not ``regular", in
the sense that their gradients would vanish on the stationary
surface. A theory with equation of motion
$\delta {\cal L}'/\delta \phi = \phi^2 = 0$ (rather than
the equivalent equation $\delta {\cal L}/
\delta \phi = \phi = 0$) would provide
such an example.
This case does not arise in usual gauge theories,
as we have just seen, but does occur in, say, Siegel formulation of
chiral bosons \cite{Siegel}.
Again, a lot of work already exists
on this subject, especially in the Hamiltonian context.  The
algebraic framework is well developed.
The real question is, however, what is the
physical meaning of the BRST construction in those
cases.  The relation between the
BRST cohomology and the cohomology of the
geometrical longitudinal derivative on the
stationary surface may no longer hold (this is why the
BRST analysis performed in chapters 9 and 10 of \cite{HT}
excludes these somewhat pathological cases).  To the author,
the question has not been fully resolved.

\section{Acknowledgements}

The author is grateful to G. Barnich,
P. Gr\'egoire, T. Kimura and
J. Stasheff for discussions.  This work has been supported in
part by a research grant from F.N.R.S. and a research contract
with the Commission of the European Community.

\end{document}